\documentclass[11pt,letterpaper]{article}
\usepackage[top=0.85in,left=2.75in,footskip=0.75in]{geometry}

\usepackage{amsmath,amssymb}

\usepackage{changepage}

\usepackage{textcomp,marvosym}

\usepackage{cite}

\usepackage{nameref,hyperref}

\usepackage[right]{lineno}

\usepackage[nopatch=eqnum]{microtype}
\DisableLigatures[f]{encoding = *, family = * }

\usepackage[table]{xcolor}

\usepackage{array}

\usepackage{mathtools}

\newcolumntype{+}{!{\vrule width 2pt}}

\newlength\savedwidth

\raggedright
\usepackage[aboveskip=1pt,labelfont=bf,labelsep=period,justification=raggedright,singlelinecheck=off]{caption}

\makeatletter
\renewcommand{\@biblabel}[1]{\quad#1.}
\makeatother

\usepackage{lastpage,fancyhdr,graphicx}
\usepackage{epstopdf}
\fancyheadoffset[L]{2.25in}
\fancyfootoffset[L]{2.25in}
\newcommand\HLB[1]{{\color{black}#1}}
\usepackage{bm}

\begin{document}
\vspace*{0.2in}

\begin{flushleft}
{\Large
\textbf\newline{Extreme value statistics of nerve transmission delay} % Please use "sentence case" for title and headings (capitalize only the first word in a title (or heading), the first word in a subtitle (or subheading), and any proper nouns).
}
\newline
\\
Satori Tsuzuki \textsuperscript{1*}
\\
\bigskip
\textbf{1} Research Center for Advanced Science and Technology, The University of Tokyo, 4-6-1, Komaba, Meguro-ku, Tokyo 153-8904, Japan
\\
\bigskip

% Use the asterisk to denote corresponding authorship and provide email address in note below.
*tsuzukisatori@g.ecc.u-tokyo.ac.jp

\end{flushleft}
\section*{Abstract}
Delays in nerve transmission are an important topic in the field of neuroscience. Spike signals fired or received by the dendrites of a neuron travel from the axon to a presynaptic cell. The spike signal then triggers a chemical reaction at the synapse, wherein a presynaptic cell transfers neurotransmitters to the postsynaptic cell, regenerates  electrical signals via a chemical reaction through ion channels, and transmits them to neighboring neurons. In the context of describing the complex physiological reaction process as a stochastic process, this study aimed to show that the distribution of the maximum time interval of spike signals follows extreme-order statistics. By considering the statistical variance in the time constant of the leaky Integrate-and-Fire model, a deterministic time evolution model for spike signals, we enabled randomness in the time interval of the spike signals. When the time constant follows an exponential distribution function, the time interval of the spike signal also follows an exponential distribution. In this case, our theory and simulations confirmed that the histogram of the maximum time interval follows the Gumbel distribution, one of the three forms of extreme-value statistics. We further confirmed that the histogram of the maximum time interval followed a Fr\'{e}chet distribution when the time interval of the spike signal followed a Pareto distribution. These findings confirm that nerve transmission delay can be described using extreme value statistics and can therefore be used as a new indicator of transmission delay.
%\linenumbers

\section*{Introduction}
\begin{figure}[t]
\vspace{-39.0cm}
\includegraphics[width=4.0\textwidth, clip, bb= 0 0 2728 2502]{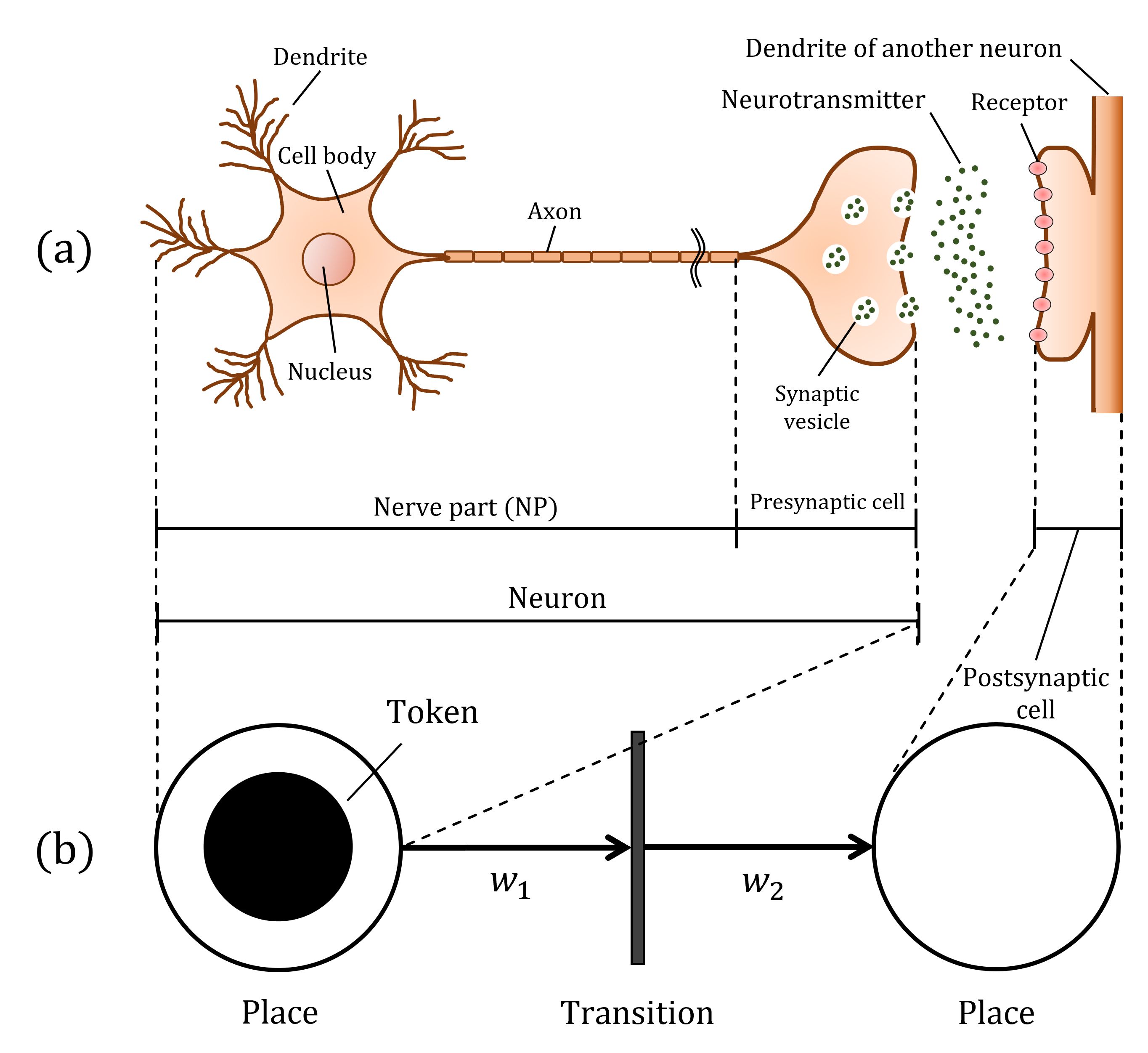}
\caption{Schematic representation of a systematic description of a single neuron. 
\HLB{(a) illustrates a single neuron and (b) represents a Petri net model graphically describing a state transition system. 
The first place in (b) corresponds to part of a neuron comprising an NP and a presynaptic cell in (a), and the second place in (b) corresponds to a postsynaptic cell in (a), with unidirectional transmissions from the presynaptic to the postsynaptic cell. 
A comparison of (a) and (b) shows that we can interpret NTU as a two-state transition problem.}}
\label{fig:Figure1}
\end{figure}

The neurological system connects multiple neurons. Spike signaling between two neurons is a complex physiological process.
A  schematic representation of a single neuron is shown in Fig.~\ref{fig:Figure1}(a), and a summary of the nerve transmission mechanism has been provided previously~\cite{Pereda2014, henley2021foundations, Xiang2021}. Each neuron has a physiological parallel resistor--capacitor circuit, in which K+ and Na+ ion channels and K+ leak channels maintain the membrane potential at the resting potential in a non-stimulated state. However, an influx of external current causes the depolarization of neurons, producing a sudden and sharp increase in potential, termed a spike. During the spike, the membrane potential exceeds a threshold value, which is followed by repolarization and hyperpolarization, and subsequently a period of preparation for the next spike generation. During this time, the Na+ ion channels enter a temporally inactive state termed the refractory period, which prevents neurotransmission backflow and ensures that the generated spike signal is transmitted in one direction in the neuron. The spike signal is then transmitted to the synapse, that is, \HLB{a structure that allows a neuron to communicate with another neuron}. Each synapse is composed of a pair of separate presynaptic and postsynaptic cells. Transmission begins with the release and diffusion of neurotransmitters into the synaptic cleft in the presynaptic cell, followed by the reception of the transmitters by receptors in the membranes of the postsynaptic cell, and subsequently by ion channels in the postsynaptic membrane that convert the transmitters into electrical signals that can then be transmitted to neighboring neurons. In this way, a neural transmission between two neurons (hereafter referred to as a ``nerve transmission unit'', or ``NTU'') occurs in a finite amount of time. The time required by an NTU fluctuates around its average value because of the physiological and chemical reactions of several minute particles, such as the transport of K+ and Na+ ions through the membrane or numerous neurotransmitters at the synapse. Thus, the time required for the signal transmission at the NTU can be described as a statistical variable rather than a deterministic constant~\cite{doi:10.1098/rspb.1965.0016, Wang1994, Sauer2016, Dutta2022}. \HLB{In the view of queueing theory}, the time between two spike signals corresponds to a waiting time in the transmission processes of the entire nervous system called a chain set of NTUs. Hereinafter, we refer to this waiting time as the ``nerve transmission period'' or ''NTP,'' defined as the total transmission time taken by a spike signal to pass through the dendrites and axons (the nerve part or NP) and the ``synaptic delay'' as defined previously literature~\cite{doi:10.1098/rspb.1965.0016}. 

To date, several studies have reported that a delay or disruption in spike signals can cause diseases such as schizophrenia~\cite{Shin2011-rz, Becker2020.08.30.273995}. Nevertheless, investigations regarding the detailed relationship between NTP and these nerve diseases are limited. Elucidating the statistical behavior of NTUs may be one way to gain deeper insight into the behavioral aspects of NTUs. Estimating the frequency distribution of the maximum delays and the upper limits from a set of time-series data for NTPs could further allow the metrics and scale parameters of the distribution to be developed as new indicators for evaluating the statistical properties of the nervous system. The impact of a disease on the nervous system can further be quantitatively evaluated by comparing the degree to which statistical metrics change from their normal state before and after the disease. Observation of NTPs over several days can help derive a theoretical histogram, such that the maximum distribution of NTPs is asymptotic, while the parameters of the theoretical histogram can be used as a new mathematical tool for evaluating neurological dysfunction.

\HLB{Historically,} the mechanism of action of NTUs has been studied by both modeling and experimentation. Hodgkin and Huxley presented deterministic time-evolution equations for an equivalent circuit model describing membrane potential changes in a neuron~\cite{hodgkin1952quantitative, ARMSTRONG1973, Aldrich1983, 10.1085/jgp.84.4.505, Clancy1999, Schwiening2012-dr}. We will refer to this model as the Hodgkin--Huxley model, or HHM. The HHM describes the detailed mechanism of the NTU as a physiological parallel resistor--capacitor circuit, as mentioned at the beginning of this section. The Leaky Integrate-and-Fire model (LIFM)~\cite{Stein2008-na, FENG2001955, Gerstner2002, Lansky2006, 10.3389/fninf.2019.00071, IGARASHI2011950} is a simplified version of the HHM that focuses on the timing and frequency of spike generation, without considering detailed generation processes, to reduce the number of variables. LIFM is more computationally friendly than HHM, and is well-suited for \HLB{simulating large-scale systems}~\cite{10.3389/fninf.2019.00071, IGARASHI2011950}. HHM or LIFM can be combined with the exponential synapse models~\cite{10.1162/neco.1993.5.2.200, doi:10.1073/pnas.1821227116, 10.3389/neuro.10.026.2009} that express a decay of the electric current at the synapse, representing the behavior of the synaptic delays. 

\HLB{In terms of stochastic processes, NTU can be interpreted as a two-state transition problem, while} the NTP can be formulated as a random variable. \HLB{The detailed reason for this is explained later using the Petri net model~\cite{brauer2009carl, petri1966communication, 10.1007/978-3-319-07674-4_51} in the Methods section. In this case}, the discrete elapsed time $T_{i}$ $(i = 1, 2, \cdots, n)$ (used as a random variable sequence) is defined as the time at which a spike signal transmission is converted into an electrical signal. \HLB{In addition}, the random variable sequence of the time difference is defined as $T_{i}$ as $d_{i} \coloneqq T_{i}-T_{i-1}$ $(i = 1, 2, \cdots, n)$, and $d_{\rm max}$ is defined as the maximum of $d_{i}$ $(i = 1, 2, \cdots, n)$ for each measurement. \HLB{Notably}, for sufficiently large $m$ measurements, a probability density distribution, $D$, can be obtained using extreme value theory (EVT)~\cite{Fisher_Tippett_1928, Frechet_Maurice_Sur_1928, Gnedenko, Taylor1985}, to which the histogram (frequency distribution) of $d_{max}$ asymptotes. EVT began with a theoretical study of the statistical distribution followed by the maximum values, which were later established with the Trinity Theorem, which shows that the distribution following such a maximum, if it exists, belongs to only one attraction domain of the three distributions: the Gumbel, Fr\'{e}chet, and Weibull distributions~\cite{Fisher_Tippett_1928, Frechet_Maurice_Sur_1928, Gnedenko, Taylor1985}. 
\HLB{Therefore, if $D$ is analytically determined using EVT}, $D$ can be used as a new indicator for objectively evaluating the maximum distribution of NTPs. \HLB{Importantly}, if a random variable sequence $X_{i}$ $(i=1,2,\cdots, n)$ follows an exponential or Pareto distribution, the distribution of the maximum order statistic of $X_{i}$ is asymptotic to the Gumbel or Fr\'{e}chet distribution of EVT, respectively~\cite{Fisher_Tippett_1928, Frechet_Maurice_Sur_1928, Gnedenko, Taylor1985} (please refer to the Methods section for a mathematical description of EVT). 
\HLB{Intriguingly, the transmission in a single NTU can be estimated to follow a Poisson distribution.} Several experiments have shown that the frequency distribution of synaptic delays follows an exponential function, strongly supporting the idea that $T_{i}$ follows a Poisson process, and that a sequence of random variables for that time interval $d_{i}$ follows an exponential distribution~\cite{doi:10.1152/jn.1999.82.3.1352, doi:10.1073/pnas.1209798109, Saveliev2015}. It follows that the asymptotic distribution of the maximum order statistic of the random variable of the time interval $d_{i}$ can be determined if $d_{i}$ follows an exponential or Pareto distribution. As mentioned above, several experiments have confirmed that NTPs follow for an exponential distribution~\cite{doi:10.1152/jn.1999.82.3.1352, doi:10.1073/pnas.1209798109, Saveliev2015}. Therefore, the distribution of the maximum values of $d_{i}$ obtained by solving the time evolution of HHM or LIFM is expected to follow the Gumbel distribution. 

\HLB{This study demonstrates the existence of an extreme value distribution of nerve transmission delays using theory and a single-neuron simulation.} To the best of our knowledge, this is the first study to apply EVT to the problem of delays in neural spike signals. \HLB{Prior research has recognized that delay occurs in the transmission of neural signals, resulting in physiological effects such as synaptic plasticity~\cite{10.1371/journal.pone.0096485, 10.1172/JCI169064, 10.1371/journal.pcbi.1002836}. However, no studies have yet elucidated the theoretical distribution of the maximum number of cases of nerve transmission delays.} 
\HLB{In particular}, we applied EVT to the problem of delays in neural spike signals that follow a Poisson process in the time direction. \HLB{For applications}, once the asymptotic distribution of the histogram of the maximum value of NTPs in nerve transmission is identified, the upper bounds and their probability of occurrence can be predicted; this could be helpful in clinical practice. Herein, we can use electroencephalography (EEG) to measure the delays in the NTPs, and EVT to estimate the upper bounds of their distribution. \HLB{ This could} serve as a stepping stone for the development of preventive measures. \HLB{In addition}, extreme value statistics \HLB{also proves that} the general extreme value distribution (GEVD)~\cite{60304975-67e2-38df-86b7-1fdac19c6683, Singh1998, Abdulali2022} \HLB{contains} these three distribution types. As GEVD is adaptable and suitable for fitting measurement data, studies that use GEVD to analyze real data are becoming popular. Semi-empirical approaches that apply GEVD to measured distributions and estimate the model parameters of GEVD using probabilistic maximum likelihood or similar methods have dominated research on spatiotemporal data~\cite{AMIN2021123, 7408339, 7500642, doi:10.1177/0954409713481725}. In particular, in areas related to neuroscience, one study reported the \HLB{ analysis } of experimental data on the velocities of dynein and kinesin in neurons with GEVD~\cite{NAOI2022400a}. \HLB{In contrast to} these data analysis studies, our study aimed to demonstrate that the histogram of maximum time intervals of neural spike signals follows EVT, both theoretically and by simulation.

\HLB{From an interdisciplinary perspective}, EVT has been used in various fields, such as in predicting the maximum amount of precipitation~\cite{sbbd_estendido, Artha_2019, doi:10.1080/23737484.2020.1789901}, the maximum age limit in a region, and the maximum speed at which an athlete can run a marathon~\cite{einmahl2008records, Einmahl2011, ArderiudeFondeville}. However, there have been few examples of the application of EVT to traffic transportation systems, even when we examine other related fields. In the fields of physics and aeronautics, we previously considered the spot assignment problem in airport ground traffic from a physical perspective ~\cite{PhysRevE.98.042102}. Focusing on a roadway consisting of a single lane with a parking lane, we solved the problem of a vehicle moving along the roadway in one direction, stopping at one of the randomly selected spots in an adjacent parking lane, and starting again after a certain period of time, using a stochastic model known as the total asymmetric simple exclusive process (TASEP)~\cite{PhysRevE.80.051119, PhysRevE.84.051127, PhysRevE.83.051128}. In our previous \HLB{work}~\cite{PhysRevE.98.042102}, we examined the stochastic process of a vehicle stopping at a spot in a parking lane while moving along a road. We found that this scenario follows the so-called generation-death process, and that it is necessary to consider the order statistics of spot usage in order to obtain the frequency distribution of spot usage. Accordingly, the spot usage distribution was described using an extreme-value statistical distribution~\cite{PhysRevE.98.042102}. This is an example of the application of EVT to stochastic processes in the \HLB{asymmetric} spatial direction. \HLB{By contrast, this study applied EVT to the problem of nerve transmission delays in the chronological direction.}

HHM and LIFM are deterministic time evolution models; however, the actual spike signal time interval was confirmed to follow an exponential distribution. As such, it is necessary to modify the deterministic time evolution models into stochastic time evolution models. This study \HLB{reproduces} the stochastic properties in the LIFM by providing a stochastic fluctuation to the time constant $\tau$, which determines the decay rate of the spike signal (see the Methods section). Briefly, each time a spike occurs, $\tau$ is updated based on a certain probability distribution. The time dependence of the voltage drop slightly fluctuates with each signal in real neurotransmission processes, making the time constant a random variable, and thus a more accurate representation of real physical phenomena. Therefore, this is a reasonable method for converting a deterministic time evolution model into a stochastic one. In summary, we demonstrated our theory that NTPs follow extreme-order statistics in a nerve transmission simulation for a single NTU using a stochastic LIFM combined with an exponential synapse model, \HLB{ which is also an emphasis of this study}. 

\section*{Methods}
\paragraph*{\bf Single neuron models.} 
The first detailed mathematical model of the transmission process of a single neuron was the HHM, formulated in 1952 by Hodgkin and Huxley, based on the experimental results of action potentials in squid axons~\cite{hodgkin1952quantitative, ARMSTRONG1973, Aldrich1983, 10.1085/jgp.84.4.505, Clancy1999, Schwiening2012-dr}. In that study, each neuron was shown to be a parallel resistance--capacitor circuit comprising K+ ion channels, Na+ ion channels, and K+ leak channels, while the spike signal was shown to propagate with a finite width waveform as a solution to the time-dependent differential equation of the electrical circuit. The HHM model accurately reproduced the neurotransmission process, including the waveform of the spike signal. However, the HHM model has some practical drawbacks, such as the fact that it retains four variables per neuron and requires a relatively accurate computational method, owing to the complexity of the equations. In physiological experiments, often only the timing of the spike signal is of interest, not the waveform of the signal. Accordingly, LIFM is an intrinsic model that simplifies the HHM by focusing only on the membrane potential shift and reducing the number of variables to one. The time evolution equation for the membrane potential in LIFM is as follows~\cite{Stein2008-na, FENG2001955, Gerstner2002, Lansky2006, 10.3389/fninf.2019.00071, IGARASHI2011950}:
\begin{eqnarray}
\tau \frac{d}{dt} V(t) = - (V(t) &-& V_{\rm rest}) + RI_{\rm ext}, \nonumber \\
V(t) > \theta~\Rightarrow~s(t) &=& 1,~V(t) \leftarrow V_{\rm reset}, \nonumber \\
	V(0) &=& V_{\rm init}. \label{eq:LIFM}
\end{eqnarray}
where $\tau$ is the time constant, $V_{\rm rest}$ is the resting potential, $V(t)$ is the membrane potential at time $t$, $R$ is the membrane resistance, $I_{\rm ext}$ is a constant external current, and $\theta$ is the threshold for spike firing. $V_{\rm reset}$ is the reset potential and $V_{\rm init}$ is the initial value of the membrane potential. The membrane potential is the equilibrium point at $V_{\rm rest}$, which asymptotes to $V_{\rm rest}$ + $R l_{\rm ext}$ at the rate of $\tau$. In addition, $s(t)$ is a discrete function equal to $1$ if a spike is fired at time $t$, but is zero otherwise. When the membrane potential $V$ exceeded $\theta$, we set $s(t) = 1$ to express that the spike was fired at that time while simultaneously resetting the membrane potential $V$ to $V_{\rm reset}$. According to Eq.~(\ref{eq:LIFM}), the change and resetting of the membrane potential are repeated and the spike signal travels through the neuron as a periodic signal.

In contrast to the original LIFM expressed in Eq.~(\ref{eq:LIFM}), Eq.~(\ref{eq:LIFM}), a deterministic differential equation, into a stochastic differential equation by replacing the time constant $\tau$ with $\xi(\mu, \sigma)$ where $\xi(\mu, \sigma)$ is the independent identity distribution given by the two parameters $\tau$ and $\sigma$, corresponding to the location and rate parameters for an exponential distribution, and the shape and location parameters for the Pareto distribution, respectively. We computed the variation in the membrane potential according to Eq.~(\ref{eq:LIFM}) after replacing the time constant $\tau$ with $\xi(\mu, \sigma)$: We assumed that the current spike signal was transmitted to the presynaptic cell when the threshold $\theta$ was reached, and then calculated the change in current at the synapse using an exponential decay model, expressed as $I_{s}(t) = I_{s}(0)e^{-t/\tau_{s}}$, where $I_s(t)$ is the current at the synapse at time $t$ and $\tau_{s}$ indicates the time required for the chemical reaction process at the synapse, that is, the synaptic delay, the membrane potential is reset when both conditions are met: the membrane potential exceeds the threshold $\theta$ and the synaptic decay is below a time constant. Therefore, the condition $V (t) > \theta$ for resetting the membrane potential in Eq. (\ref{eq:LIFM}) is modified as $V (t) > \theta \cap I_{s}(t) < \omega$, where $\omega$ corresponds to $I_{s}(0)e^{-1}$. Accordingly, we calculated both the spike signal transmission and synaptic delay, which makes the description closer to the actual nerve transmission compared to Eq.~(\ref{eq:LIFM}). Nevertheless, in practice, the effect of synaptic delay is negligible in a single-neuron simulation in one direction. This can be explained as follows: In the case of multiple neurons, the current $I_{s}$ at the synapse feeds back to the variation in the membrane potential as $I_{\rm ext}$ in some cases. However, in a single-neuron problem, the signal flows in one direction. In addition, it decays immediately compared to the variation in the membrane voltage. Therefore, $I_{s}$ does not affect the behavior of the entire system in our case because we focus on the distribution of the maximum time intervals when the synaptic time intervals follow an exponential distribution. In summary, modified versions of Eq. (\ref{eq:LIFM}) can be expressed as follows.
\begin{eqnarray}
\xi(\mu, \sigma) \frac{d}{dt} V(t) = - (V(t) &-& V_{\rm rest}) + RI_{\rm ext}, \nonumber \\
V(t) > \theta \cap I(t) < \omega~\Rightarrow~s(t) = &1&, V(t) \leftarrow V_{\rm reset}, \nonumber \\
V(0) &=& V_{\rm init}.  
\label{eq:LIFMmodif}
\end{eqnarray}

\paragraph*{\bf \HLB{Assumption of the Poisson process.}}
Neural transmission in an NTU can be expressed using a directed graph consisting of two nodes. For clarity, we explain the NT system using the classical Petri net model~\cite{brauer2009carl, petri1966communication, 10.1007/978-3-319-07674-4_51}, which is an established representation of state-transition diagrams. We outline this Petri net model below, with reference to Fig.~\ref{fig:Figure1}(b). First, the components of a system or their states are depicted by a circular symbol called ``place'' and an event is depicted by a bar-shaped symbol called ``transition.'' The correlation between the components of place and transition can be expressed by a directed arrow called ``arc.'' The objects processed on the system are represented by a black filled circle called a ``token.''  The flow of objects was traced on a Petri net diagram using these tokens. The first place in (b) corresponds to a part of a neuron consisting of an NP and a presynaptic cell in (a), and the second place in (b) corresponds to a postsynaptic cell in (a), with unidirectional transmission from the presynaptic to the postsynaptic cell. Additionally, the ``transition'' shows all the events, including the spike generation, release, diffusion, reception, and conversion of neurotransmitters into electrical signals. The ``token'' shown as a black circle in (b) indicates the spike signals flow at the NP or neurotransmitters at the synapse. The weights $w_{1}$ and $w_{2}$ represent the conditions that can fire the transition in each arc; these weights usually represent the firing criteria. When the number of tokens exceeds the weight, the tokens are ready to pass through the arc. In NT problems, we regard arcs as always transmittable states as long as we ignore exceptional dysfunction phenomena such as synaptic fatigue~\cite{Simons-Weidenmaier2006}, where neurotransmitters in the synapse become scarce and transmission is aborted. Hence, we can ignore the conditions for $w_{1}$ and $w_{2}$ in the first approximation and set them to 1. A comparison of (a) and (b) clearly shows that we can interpret NTU as a two-state transition problem.

The transmission in a single NTU can be assumed to follow a Poisson distribution for the following reasons: Considering a single NTU in a nerve chain connecting multiple NTUs which is sufficiently long to ignore the effect of the edges of the chain, we can assume an open boundary condition. Due to the refractory period, the postsynaptic cells in the NTU did not receive other signals during each signal transmission. Therefore, we can assume each event to be independent, where each ``event'' is defined as the transmission process from the generation of a spike signal at a neuron to its reception by receptors, as well as its conversion into electrical signals at a postsynaptic cell. As a thought experiment, we consider observing the transmission process in the NTU for finite time $T$. Owing to physiological homeostasis, it is reasonable to consider the transmission flow in a steady state. In addition, in a zero-order approximation, the transmission rate can be assumed to be invariant during the observation. The invariant transmission rate and independence of each event imply that the probability of events occurring at any time interval $(t_{i}, t_{i} + t)$ is independent of the events that occurred before $T_{i}$. Therefore, the system is memory-less. The reason for stating ``in the zero-order approximation'', is that at this stage, we ignore synaptic fatigue~\cite{Simons-Weidenmaier2006}, where neurotransmitters in the synapse become scarce and transmission is aborted. Consider discrete time $\Delta t$ as a short time of the same order as $d_{i}$. The probability of an event occurring more than once during $\Delta t$ is negligible, because $d_{i}$ is the interval in which an event occurs. Accordingly, this system exhibited event scarcity. Thus, the system has the following conditions to be considered a Poisson distribution: stationarity, independence, memorylessness, and scarcity. Thus, $T_{i}$ is a sequence of random variables following a Poisson process. Recall that $d_{i}$ is a sequence of random variables for the time interval $T_{i}-T_{i-1}$ $(i=1,2,\cdots, n)$. If a sequence of random variables follows a Poisson process, then their difference sequence of random variables follows an exponential distribution~\cite{cinlar2013introduction}. Therefore, $d_{i}$ is expected to follow the exponential distribution. \HLB{As mentioned above}, several experiments have shown that the frequency distribution of synaptic delays follows an exponential function, strongly supporting the idea that $T_{i}$ follows a Poisson process and that a sequence of random variables for that time interval $d_{i}$ follows an exponential distribution~\cite{doi:10.1152/jn.1999.82.3.1352, doi:10.1073/pnas.1209798109, Saveliev2015}.

If a random variable sequence $X_{i}$ $(i=1,2,\cdots, n)$ follows an exponential or Pareto distribution, then the distribution of the maximum order statistic of $X_{i}$ is asymptotic to the Gumbel or Fr\'{e}chet distribution of EVT, respectively~\cite{Fisher_Tippett_1928, Frechet_Maurice_Sur_1928, Gnedenko, Taylor1985} (see the \HLB{below} for the mathematical description of EVT). It subsequently follows that the asymptotic distribution of the maximum order statistic of the random variable of the time interval $d_{i}$ can be determined if $d_{i}$ follows an exponential or Pareto distribution. Since NTPs can be assumed to follow an exponential distribution~\cite{doi:10.1152/jn.1999.82.3.1352, doi:10.1073/pnas.1209798109, Saveliev2015}, the distribution of the maximum values of $d_{i}$ obtained by solving the time evolution of HHM or LIFM is expected to follow the Gumbel distribution. 

\begin{figure}[t]
\vspace{-45.0cm}
\begin{center}
\includegraphics[width=4.1\textwidth, clip, bb= 0 0 2542 2584]{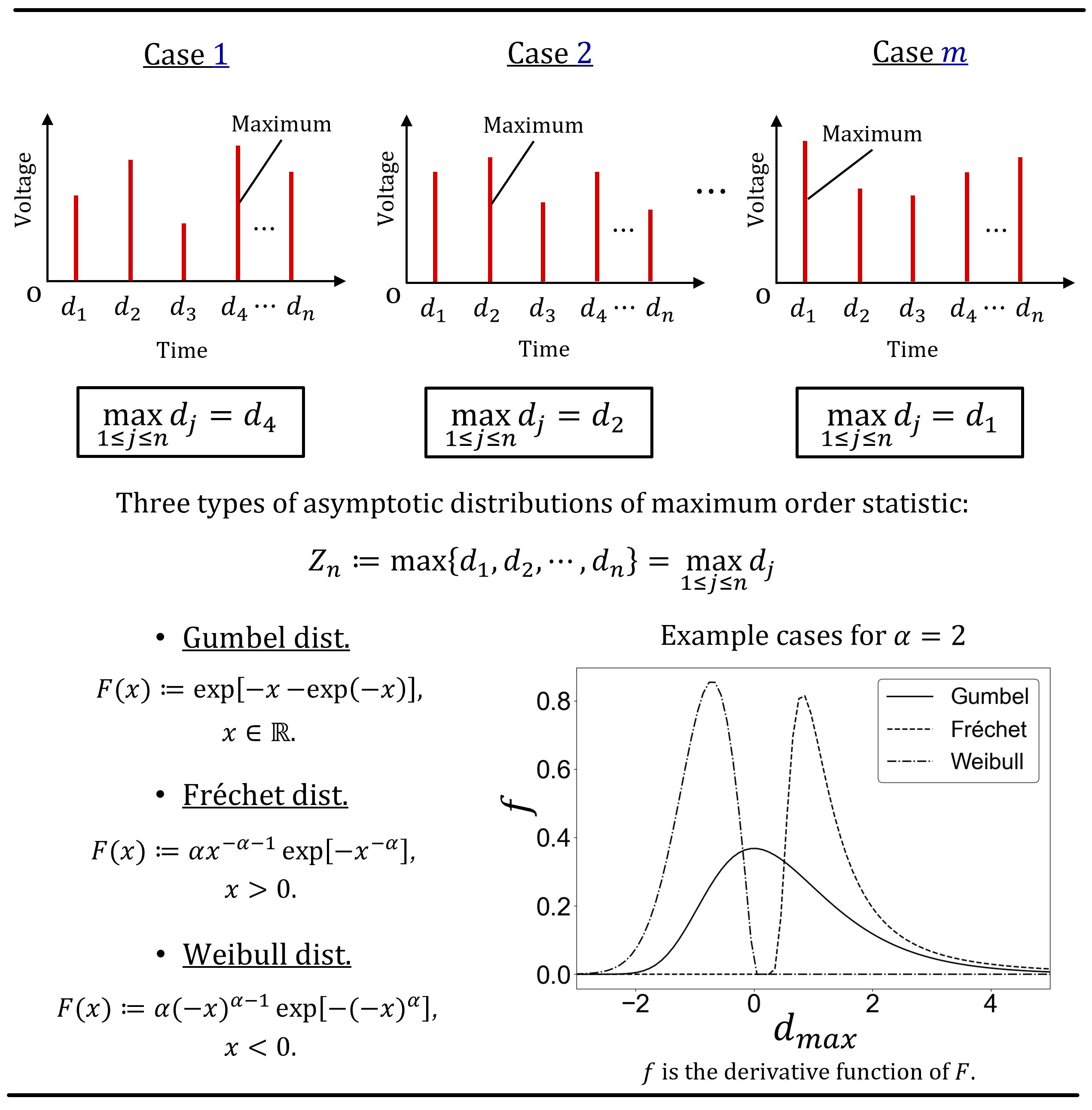}
\end{center}
\caption{Schematic view of the extreme order statistics. \HLB{The upper part of the figure schematically describes the maximum order statistic when $m$ data sets each of which consists of $n$ data are measured. The maximum order statistic is the statistic of the maximum values sampled from each of the sufficient $m$ datasets, each of which comprises sufficiently large $n$ data. Extreme Value Theory (EVT) examines the asymptotic distribution of the maximum order statistic, revealing that there are three types of extreme value distributions: Gumbel, Fr\'{e}chet, and Weibull distributions.}}
\label{fig:Figure2}
\end{figure}

\paragraph*{\bf Extreme value theory (EVT).}
The maximum order statistic of the independent identical random variable is defined as $X_{j}~(j=1,2,\cdots,n)$ as $Z_{n}$. Mathematically, this definition can be expressed as $Z_{n}$~$\coloneqq$~${\rm max}\{X_{1}, X_{2},\cdots, X_{n}\}$~$=$~${\rm max}~X_{j}~(j=1,2,\cdots, n)$. Under this definition, EVT discusses the asymptotic distribution of $Z_{n}$ to obtain sufficient measurements of $Z_{n}$. Figure~\ref{fig:Figure2} schematically explains the details of the extreme-value statistics. Suppose we measured $m$ datasets of $Z_{n}$ in the example shown in Fig. ~\ref{fig:Figure2}, $Z_{n}$ is $d_{4}$ for Case 1, $d_{2}$ for Case 2, and $d_{1}$ for Case $m$. If $m$ is sufficiently large, we can obtain the histogram (frequency distribution) of $Z_{n}$. In particular, if we find $k$ counts of data such that $Z_{n}$ becomes a value of $Y$, then there are $_{m} C _{k}$ possible combinations in which we can observe $Y$ for $k$ times among $m$ measurements. It is necessary to consider all of these points in order to derive the probability density function (PDF) for $Z_{n}$. The PDF can be obtained as the first derivative of the cumulative distribution function (CDF), EVT begins by deriving the CDF of $Z_{n}$. Now, consider the case in which the independent identical variable $X_{i}$ has a population distribution $F$. The CDF of $Z_{n}$ is expressed as $P(Z_{n} \le x)$, where $P$ is the probability and $x$ is a continuous point. From the definition of $Z_{n}$, we can derive the following relationship. 
\begin{eqnarray}
P(Z_{n} \le x)~&=&~P\bigl(\underset{1 \le j \le n}{\rm max}~X_{j} \le x \bigr)\nonumber \\
		       &=&~P(X_{j} \le x, j=1,2,\cdots, n) \nonumber \\
		       &=&~\prod_{j=1}^n P(X_{j} \le x) \nonumber \\
			   &=&~F^{n}(x). \label{eq:derivemultiple} 
\end{eqnarray}
Equation (\ref{eq:derivemultiple}) shows that the CDF of $Z_{n}$ is equal to the population distribution $F$ to the power of $n$. Thus, we focus on the distribution where $F^{n}(x)$ converges. Fisher and Tippett~\cite{Fisher_Tippett_1928} mathematically already proved the following equivalence relationship with respect to the convergence of $F^{n}(x)$:
\begin{eqnarray}
 F \in \mathcal{D}(G) \Leftrightarrow \lim_{n \to \infty}F^{n}(a_{n}x+b_{n}) &=& G(x), \nonumber \\
	 a_{n} > 0, b_{n} \in \mathbb{R}. \label{eq:extstat:cond} 
\end{eqnarray}
Here, $G$ is a continuous distribution that is neither degenerate nor divergent. $F \in \mathcal{D}(G)$ indicates that $F$ belongs to the attraction domain of $G$. 
Equation (\ref{eq:extstat:cond}) states the following: if $F \in \mathcal{D}(G)$, i.e., if the population distribution $F$ of independent identical random variable $X_{j}~(1 \le j \le n)$ belongs to the attraction domain of $G$, which is the asymptotic distribution of the maximum order statistic $Z_{n}$ and is neither degenerate nor divergent, $a_{n}$ and $b_{n}$ exist for which $F^{n}(a_{n} x + b_{n})$ converges in $G$ for sufficiently large $n$. Conversely, if we find a pair $a_{n}$ and $b_{n}$, where $F^{n}(a_{n} x + b_{n})$ converges to $G$, $F$ belongs to the attraction domain of $G$. Equation (\ref{eq:extstat:cond}) also implies that a linear transformation of $G$ with respect to the scale $a_{n}$ and location $b_{n}$ is permitted to avoid degeneration or divergence. This can be understood by employing $\bar{x} = a_{n} x + b_{n}$ and rewriting Eq.~(\ref{eq:extstat:cond}) for $\bar{x}$. 

The Trinity Theorem by Fisher and Tippett~\cite{Fisher_Tippett_1928}, Fr\'{e}chet~\cite{Frechet_Maurice_Sur_1928}, and Gnedenko~\cite{Gnedenko} proves that only three types of extreme distributions satisfy Eq. (\ref{eq:extstat:cond}): Gumbel, Fr\'{e}chet, and Weibull distributions. This theorem also proves that any population distribution $F$ is asymptotic to one of the three extreme distributions if $F \in \mathcal{D}(G)$. The CDFs of the Gumbel, Fr\'{e}chet, and Weibull distributions are as follows:
\begin{eqnarray}
{\rm Gumbel}: G(x) &\coloneqq& {\rm exp}[-{\rm exp}(-x)],~x\in \mathbb{R}, 
\label{eq:extstat:ttt1}\\
{\rm Fr{\rm \acute{e}}chet}: G(x) &\coloneqq& {\rm exp}(-x^{-\alpha}),~x\ge 0,~\alpha > 0, 
\label{eq:extstat:ttt2} \\
{\rm Weibull}: G(x) &\coloneqq& {\rm exp}[-(-x)^{\alpha}],~x\le 0, \alpha \ge 0. 
\label{eq:extstat:ttt3} 
\end{eqnarray}
If $X_{j}$ $(j=1,2,\cdots,n)$ follows an exponential distribution, $F$ can be expressed as $F(x)=1-{\rm exp}(-x)$,~$x\in \mathbb{R}$. In this case, the convergence of $F^{n}(x)$ for a sufficiently large $n$ can be identified by selecting $a_{n}$ and $b_{n}$ as 1 and ${\rm log}(n)$ as follows~\cite{PhysRevE.98.042102}:
\begin{eqnarray}
&~& \lim_{n \to \infty} F^{n}(a_{n}x + b_{n}) \nonumber \\
&=& \lim_{n \to \infty}\Biggl\{ 1+\frac{-n[1-F(a_{n}x+b_{n})]}{n}\Biggr\}^{n} \nonumber \\
&=& \lim_{n \to \infty}\Biggl\{ 1+\frac{-{\rm exp}(-x)}{n}\Biggr\}^{n}  \nonumber \\
&=& {\rm exp}[-{\rm exp}(-x)], ~x\in \mathbb{R}.
\label{eq:extstat:exp1}
\end{eqnarray}
Therefore, the maximum order statistic $Z_{n}$ of the random variable $X_{j}$ $(1 \le j \le n)$ converges to the Gumbel distribution if $X_{j}$ follows an exponential distribution. Meanwhile, if $X_{j}$ $(1 \le j \le n)$ follows a Pareto distribution, $F$ is expressed as $F(x) = 1-1/x^{\alpha}$, where $\alpha > 0$ and $x \ge 1$. In this case, by using a derivation similar to that in the exponential distribution, $F^{n}(a_{n}x + b_{n})$ converges to the Fr\'{e}chet distribution by selecting $a_{n}$ and $b_{n}$ as $n^{1/\alpha}$ and $0$, respectively. 
\begin{eqnarray}
\lim_{n \to \infty} F^{n}(a_{n}x + b_{n}) &=& {\rm exp}(-x^{-\alpha}). \label{eq:extstat:pareto}
\end{eqnarray}
In summary, $Z_{n}$ converges to the Gumbel distribution if $X_{j}$ $(1 \le j \le n)$ follows an exponential distribution, and the Fr\'{e}chet distribution if $X_{j}$ $(1 \le j \le n)$ follows a Pareto distribution.

\section*{Analysis}
Stochastic LIFM simulations were performed using Eq.~(\ref{eq:LIFMmodif}) for two different cases, in which $\xi(\mu, \sigma)$ follows an exponential or Pareto distribution. In our tests, we set ($V_{\rm rest}, V_{\rm reset}, \theta, R, I_{\rm ext}$) to be (-65 mV, -65 mV, -55 mV, 1.0 MOhm, 12 nA) by reference to the literature~\cite{Sekirnjak2002, doi:10.1152/jn.00510.2004, doi:10.1152/jn.1997.77.2.537}. We further set the small time $\Delta t$ and the number of time steps $n$ to $1$~ms and $1000$ iterative steps, respectively; $1000$ ms of physical time was computed in each simulation. We used the inverse transformation method~\cite{doi:10.1137/1.9780898717570, Lecuyer2011} to generate a time interval $d_{i}$ (${i=1,2,\cdots,n}$) random sequence of spike signals following exponential and Pareto distributions. In the inverse transformation method, the random sequence $d_{i}$ (${i=1,2,\cdots,n}$) is given by $d_{i} = \xi(\mu, \sigma) = (-1.0/\sigma){\rm log} (U) + \mu$ for exponential distributions, where $U$ denotes a uniform distribution. In this formula, $\sigma$ corresponds to the rate parameter of the exponential function that determines the distribution scale. The parameter $\mu$ shifts the distribution in the positive direction to an asymptote of $\mu$; this parameter corresponds to the parameter $\tau$ in Eq.~(\ref{eq:LIFM}). Accordingly, we set $\mu$ to $20$ ms and $\sigma$ to 5 ms. By contrast, for the Pareto distribution, the random sequence $d_{i}$ (${i=1,2,\cdots,n}$) is given by $d_i = \xi(\mu, \sigma) = \mu/{(U)}^{1.0/\sigma}$. In this model, the location and attenuation rate are given by parameters $\mu$ and $\sigma$, respectively. We set the parameter $\mu$ to 20 ms. On the other hand, we performed simulations for two different cases: $\sigma$ = 7.5 and $\sigma$ = 20.

\begin{figure}[t]
\vspace{-32.5cm}
\begin{center}
\includegraphics[width=4.4\textwidth, clip, bb= 0 0 2505 1764]{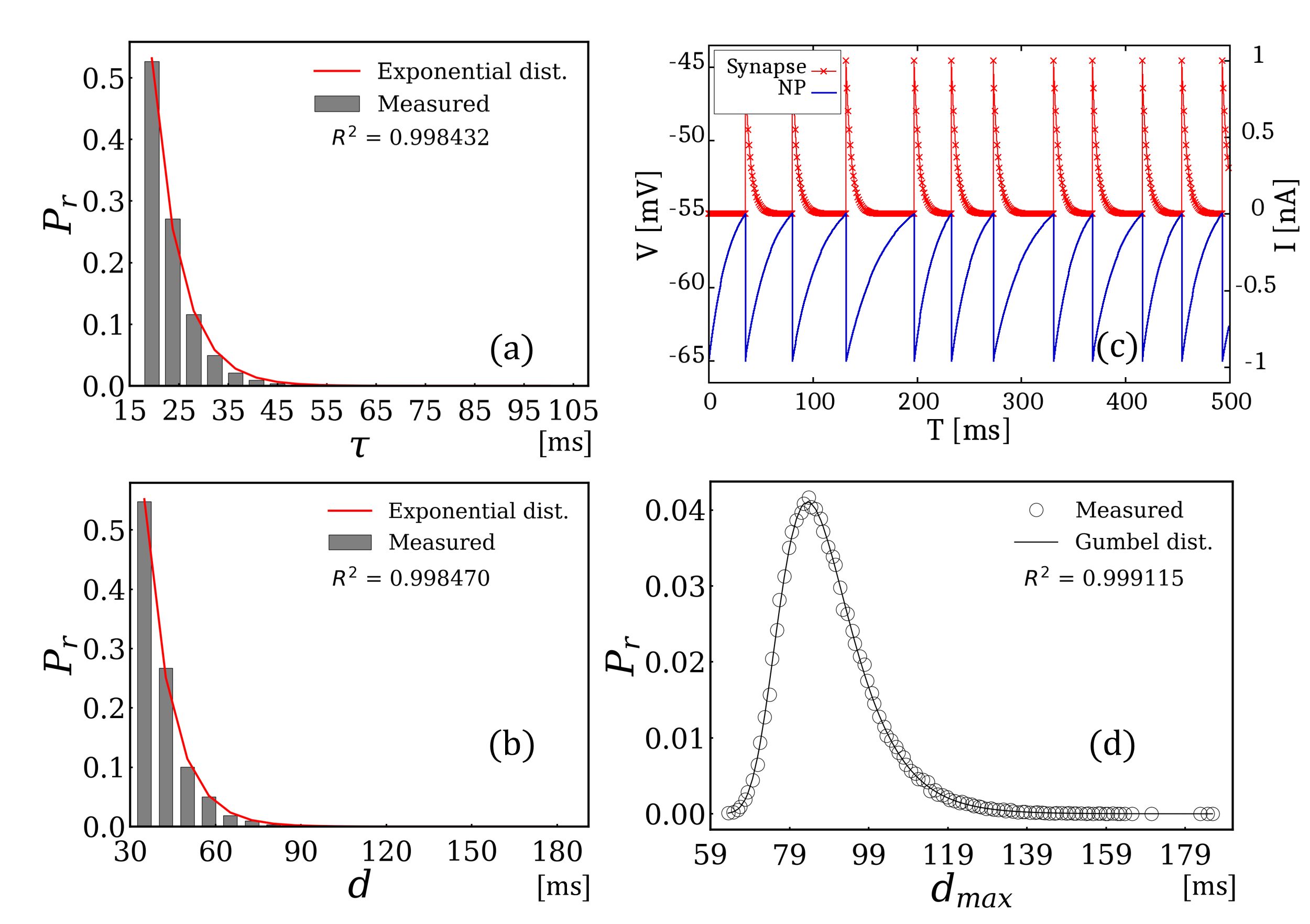}
\end{center}
\caption{Simulation results when $\xi(\mu, \sigma)$ follows an exponential distribution: (a) distribution of $\xi(\mu, \sigma)$, (b) distribution of the time interval, (c) example snapshot of spike signals, and (d) obtained Gumbel distribution.}
\label{fig:Figure3}
\end{figure}

Figure~\ref{fig:Figure3} presents the simulation results when $\xi(\mu, \sigma)$ follows an exponential distribution. First, Fig.~\ref{fig:Figure3}(a) indicates that $\xi(\mu, \sigma)$ follows an exponential distribution. Figure~\ref{fig:Figure3}(b) shows the resulting time intervals of the spike signals obtained by solving Eq.~(\ref{eq:LIFMmodif}) with a randomly updated $\xi$, when each spike signal is generated. This confirms that the time interval of the generated spike signals followed an exponential distribution. Figure~\ref{fig:Figure3}(c) shows the variation in current $I$ (which corresponds to $I_{s}$ in the Methods section) at the synapse and membrane voltages $V$. Once the voltage reached 55 mV, it triggered the generation of a spike signal, which was immediately transmitted to the synapse, where the current was attenuated using the exponential decay model. As mentioned earlier, in a single-neuron problem, the signal flows in one direction, and the current $I_{s}$ immediately decays compared to the variation in the membrane voltage. Consequently, $I_{s}$ does not affect the behavior of the entire system.  
The circular dots shown in Fig.~\ref{fig:Figure3}(d) present a histogram of the maximum time interval measured over 100000 trials. Each trial iterated 1000 time steps for the time evolution of Eq.~(\ref{eq:LIFMmodif}). The histogram was normalized such that the distribution sum was 1. Notably, the obtained histogram corroborated the Gumbel distribution, which is consistent with EVT. 
Note that for the values of R-squared, the coefficient of determination was greater than 0.99 in all cases. Accordingly, we demonstrated that in single-neuron problems, the histogram of the maximum time intervals followed a Gumbel distribution when the time intervals followed an exponential distribution.

\begin{figure}[t]
\vspace{-46.3cm}
\begin{center}
\includegraphics[width=4.5\textwidth, clip, bb= 0 0 2490 2384]{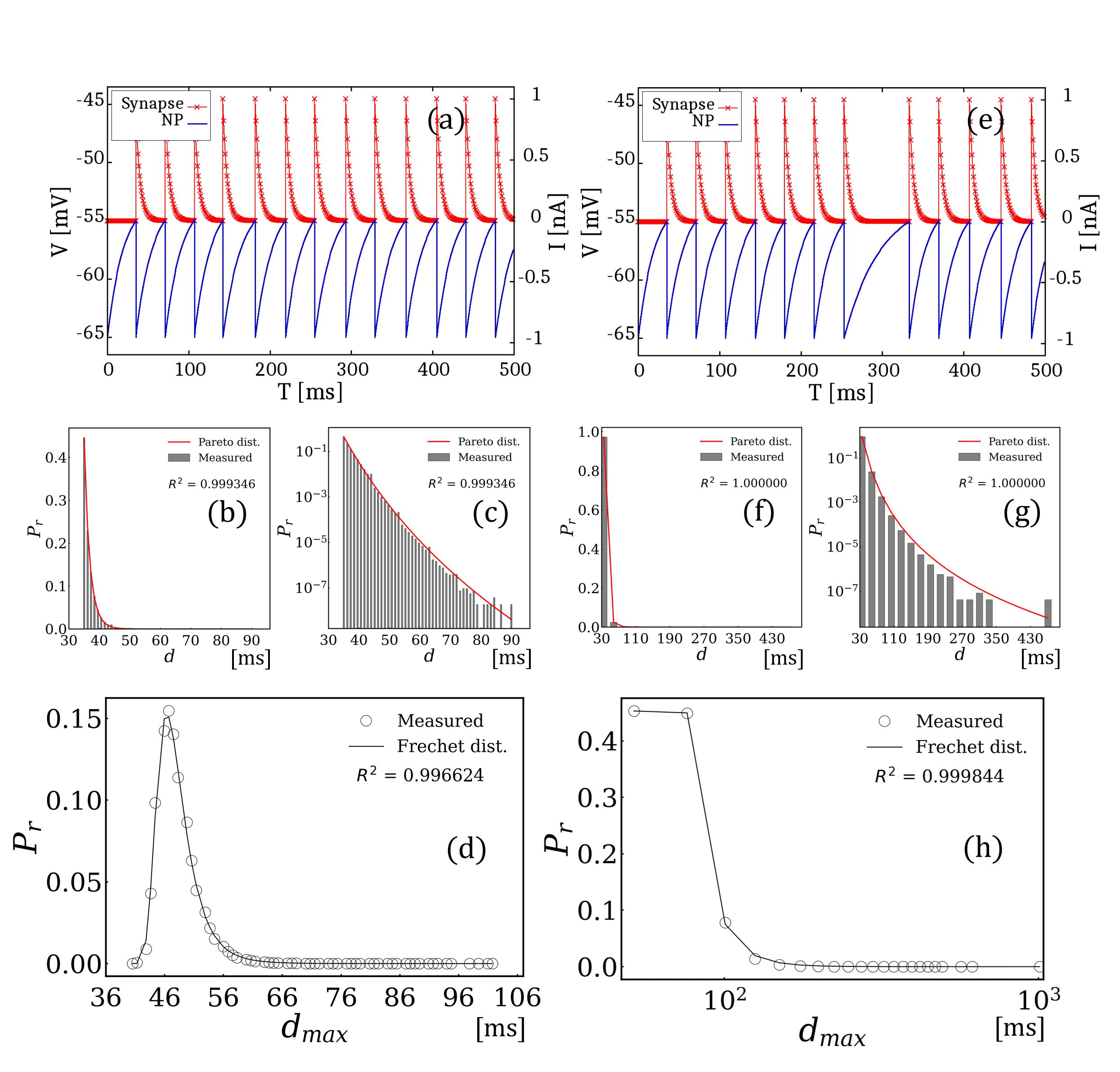}
\end{center}
\caption{Simulation results when $\xi(\mu, \sigma)$ follows a Pareto distribution: (a)–(d) shows the results for $\sigma = 20$: (a) example snapshot of spike signals, (b) distribution of the time interval, (c) logarithmic scale of (b), and (d) obtained Frechet distribution. In contrast, (e)–(h) show the results for $\sigma = 7.5$: (e) example snapshot of spike signals, (f) distribution of the time interval, (g) logarithmic scale of (f), and (h) obtained Frechet distribution.}
\label{fig:Figure4}
\end{figure}

Figure~\ref{fig:Figure4} shows the simulation results for the case in which $\xi(\mu, \sigma)$ follows a Pareto distribution in the small and large cases of $\sigma = 7.5$ or $\sigma = 20$. As reported in previous transmission delay experiments, the distribution of NTPs primarily follows exponential distributions~\cite{doi:10.1152/jn.1999.82.3.1352, doi:10.1073/pnas.1209798109, Saveliev2015}. However, in some cases, the NTP distributions exhibited sharper peaks and longer tails, similar to those of the Pareto distribution. In response to these observations, we investigated a histogram of the maximum NTPs for the Pareto distribution. Figures~\ref{fig:Figure4}(a)-(d) shows the results for $\sigma = 20$. First, (a) shows an example snapshot of the spike signals for reference, while (b) shows the obtained distribution of time intervals and fitting results using a Pareto distribution. The results indicate that the measured distribution followed a Pareto distribution. Because the maximum and minimum values are significantly separated in the Pareto distribution, we show the logarithmic scale of (b) in (c) to confirm the existence of the long tail, which is a trademark of the Pareto distribution. Figure~\ref{fig:Figure4}(d) shows a histogram of the maximum time intervals measured in one million trials. Each trial iterated 1000 time steps for the time evolution of Eq.~(\ref{eq:LIFMmodif}). The histogram was normalized such that the distribution sum was 1. The value of R-squared, which is the coefficient of determination, was greater than 0.99. Notably, the histogram obtained followed the Frechet distribution, which was consistent with EVT. 

The Figs.~\ref{fig:Figure4}(e)-(h) shows the results for $\sigma = 7.5$. Figure~\ref{fig:Figure4} (e) shows a snapshot of a spike signal. Figures~\ref{fig:Figure4} (f) and (g) shows the distribution of time intervals on normal and logarithmic scales, respectively. Unlike the case of $\sigma = 20$, a drastically delayed case of $d > 430$ ms for $\sigma = 7.5$ was observed. Even in this extreme case, a Pareto distribution fits the measured distribution of the time intervals while maintaining its characteristic high peak and long tails.  Figure~\ref{fig:Figure4}(h) shows a histogram of the maximum time intervals measured in one million trials. Each trial iterated 1000 time steps for the time evolution of Eq. (\ref{eq:LIFMmodif}). The histogram was normalized such that the distribution sum was 1. Notably, the obtained histogram followed the Frechet distribution, which is consistent with theory. Accordingly, we reported that in synaptic problems, the histogram of the maximum time intervals followed the Fr\'{e}chet distribution when the time intervals followed a Pareto distribution.

\section*{\HLB{Discussion}}
\HLB{For classical single-neuron experiments such as the aforementioned squid experiments~\cite{hodgkin1952quantitative, ARMSTRONG1973, Aldrich1983, 10.1085/jgp.84.4.505, Clancy1999, Schwiening2012-dr}, the extreme value distributions obtained in this study can directly explain the frequency distribution of the maximum delay between the arrival intervals of spike signals with statistical fluctuations. Conversely, for a more realistic neuronal series with a myriad of connected neurons, we should be reminded that S. Boudkkazi~\cite{Boudkkazi2007-lk, Boudkkazi2011-pn, Boudkkazi2024-bq} pointed out that both short- and long-term synaptic plasticity may be caused by the modulation of synaptic delay. They further suggested that the amplitude and duration of presynaptic action potentials determine the synaptic delay at excitatory synapses in the hippocampus and neocortex. In brief, they argued that the features (amplitude and duration) of the spike signal may determine synaptic plasticity. This situation is consistent with that described by the LIFM, with stochastic fluctuations in the time constants presented in this study. Therefore, the extreme-value statistical distribution, which is the frequency distribution of the maximum delay of the spike signal obtained in this study, may directly represent the frequency distribution that causes short- or long-term synaptic plasticity. However, it may be difficult to distinguish between the two because of the transition from short- to long-term synaptic plasticity. Synaptic plasticity is strongly associated with learning and memory~\cite{10.1371/journal.pone.0096485, 10.1172/JCI169064, 10.1371/journal.pcbi.1002836}. Thus, the extreme-value statistical distribution, which is the frequency distribution of the maximum latency, may be an indicator of memory capacity, or, conversely, forgetfulness, i.e., the frequency of short-term memory lapses.

On the other hand, in real neural networks, neurons are connected not only in series, but also in parallel. In a parallel-connected system, the probability of receiving a spike signal at neuron $i$ is the sum of the probabilities that a spike signal is transmitted to neuron $i$ from each of several input neurons independently connected to neuron $i$. Consider the stochastic process of an event in which the spike signal arrives at neuron $i$. The probability of receiving a spike signal from one neuron $k$ $(k=1,2,\cdots,n_{a})$ of the $n_{a}$ neurons connected to neuron $i$ can be represented as a Poisson process, according to the Poisson distribution characterized by the parameter $\lambda_{k}$. In this case, the sequence of random variables at the time of arrival of the spike signal at neuron $i$ is characterized by the parameter $\sum^{n_a}_{k=1} \lambda _k (k=1,2,\cdots, n_{a})$. Therefore, as discussed in this study, extreme-value statistical distributions can also be discussed. Nevertheless, to obtain the extreme value distribution of the maximum delay in a real neural network with a parallel system, it may be easier to fit the frequency distribution of the maximum value of the measured data in a neuron (corresponding to neuron $i$ above) using GEVD rather than deriving the distribution analytically or theoretically. Izhikevich further reported on the observation of polychronization phenomena in parallel neuronal systems ~\cite{Izhikevich2006-ll}. Several studies in related fields have reported the synchronization of events characterized by extreme-value distributions~\cite{Chowdhury_2019, 10.1063/5.0131133}. This suggests that if the frequency distribution of the maximum delay specific to each group of neurons in a parallel-connected system can be obtained using GEVD, then the synchronization phenomena of their different extreme-value distributions may also be investigated. In other words, it is possible to discuss the relationship between the synchronization patterns of different extreme value distributions and physiological phenomena such as learning, memory, and forgetting. These issues should be investigated further in future studies.
}

\section*{Conclusion}
This study presents nerve transmission, a complex physiological reaction, as a stochastic process. This study showed that the distribution of the maximum time interval of the spike signals followed extreme-order statistics. By introducing statistical variance into the time constant of the Leaky Integrate-and-Fire model, a deterministic time evolution model of spike signals, we allowed randomness to emerge in the time interval of the spike signals. We further confirmed that the time interval of the spiked signal followed an exponential distribution if the time constant followed an exponential distribution function. In this case, our theory and simulations confirmed that the histogram of the maximum time interval followed the Gumbel distribution, which is one of the three types of extreme-value statistics. In addition, we confirmed that the histogram of the maximum time interval followed a Fr\'{e}chet distribution when the time interval of the spike signal followed a Pareto distribution. Our results showed that neurotransmission delays can be described using extreme value statistics. In this study, we used the classical LIFM model. Although several improved LIFM models have been presented previously, as long as the elapsed time of the spike signal transmission follows a Poisson process, the results from the improved models with excellent waveform reproducibility follow the same patterns as those presented in this study. 
\HLB{In addition, the relationship between the extreme value distribution of spike signals and physiological phenomena, such as synaptic plasticity, memory, and forgetting was discussed, as well as the potential for applying the findings to more realistic networks.}
Overall, in this study, we successfully demonstrated that the distribution of extreme values is a new indicator of nerve transmission problems. 
%\linenumbers

%\section{}
%\label{}



%% \bibliographystyle{elsarticle-num} 
%% \bibliography{<your bibdatabase>}


%\appendix
\section*{Acknowledgment}
%This study was supported by JSPS KAKENHI, Grant Number 22K14177. 
The authors would like to thank Editage (www.editage.jp) for the English language editing.

\section*{Data availability statement}
No new data were created or analyzed in this study.

\nolinenumbers

% Either type in your references using
% \begin{thebibliography}{}
% Text
% \end{thebibliography}
%
% or
%
% Compile your BiBTeX database using our plos2015.bst
% style file and paste the contents of your .bbl file
% here. See http://journals.plos.org/plosone/s/latex for further detail. 
\bibliographystyle{plos2015}
\bibliography{reference}

\begin{thebibliography}{10}

\bibitem{Pereda2014}
Pereda AE.
\newblock Electrical synapses and their functional interactions with chemical
  synapses.
\newblock Nature Reviews Neuroscience. 2014;15(4):250--263.
\newblock doi:{10.1038/nrn3708}.

\bibitem{henley2021foundations}
Henley C.
\newblock Foundations of neuroscience.
\newblock [sn]; 2021.

\bibitem{Xiang2021}
Xiang Z, Tang C, Chang C, Liu G.
\newblock A new viewpoint and model of neural signal generation and
  transmission: Signal transmission on unmyelinated neurons.
\newblock Nano Research. 2021;14(3):590--600.
\newblock doi:{10.1007/s12274-020-3016-1}.

\bibitem{doi:10.1098/rspb.1965.0016}
Katz B, Miledi R.
\newblock The measurement of synaptic delay, and the time course of
  acetylcholine release at the neuromuscular junction.
\newblock Proceedings of the Royal Society of London Series B Biological
  Sciences. 1965;161(985):483--495.
\newblock doi:{10.1098/rspb.1965.0016}.

\bibitem{Wang1994}
Wang J, Miller MI, Ogielski AT.
\newblock In: Eeckman FH, editor. A Stochastic Model Of Synaptic Transmission
  and Auditory Nerve Discharge (Part I). Boston, MA: Springer US; 1994. p.
  147--152.
\newblock Available from: \url{https://doi.org/10.1007/978-1-4615-2714-5_24}.

\bibitem{Sauer2016}
Sauer M, Stannat W.
\newblock Reliability of signal transmission in stochastic nerve axon
  equations.
\newblock Journal of Computational Neuroscience. 2016;40(1):103--111.
\newblock doi:{10.1007/s10827-015-0586-0}.

\bibitem{Dutta2022}
Dutta S, Detorakis G, Khanna A, Grisafe B, Neftci E, Datta S.
\newblock Neural sampling machine with stochastic synapse allows brain-like
  learning and inference.
\newblock Nature Communications. 2022;13(1):2571.
\newblock doi:{10.1038/s41467-022-30305-8}.

\bibitem{Shin2011-rz}
Shin YW, O'Donnell BF, Youn S, Kwon JS.
\newblock Gamma oscillation in schizophrenia.
\newblock Psychiatry Investig. 2011;8(4):288--296.

\bibitem{Becker2020.08.30.273995}
Becker S, Nold A, Tchumatchenko T.
\newblock Formation and synaptic control of active transient working memory
  representations.
\newblock bioRxiv. 2020;doi:{10.1101/2020.08.30.273995}.

\bibitem{hodgkin1952quantitative}
Hodgkin AL, Huxley AF.
\newblock A quantitative description of membrane current and its application to
  conduction and excitation in nerve.
\newblock The Journal of physiology. 1952;117(4):500.

\bibitem{ARMSTRONG1973}
ARMSTRONG CM, BEZANILLA F.
\newblock Currents Related to Movement of the Gating Particles of the Sodium
  Channels.
\newblock Nature. 1973;242(5398):459--461.
\newblock doi:{10.1038/242459a0}.

\bibitem{Aldrich1983}
Aldrich RW, Corey DP, Stevens CF.
\newblock A reinterpretation of mammalian sodium channel gating based on single
  channel recording.
\newblock Nature. 1983;306(5942):436--441.
\newblock doi:{10.1038/306436a0}.

\bibitem{10.1085/jgp.84.4.505}
Horn R, Vandenberg CA.
\newblock {Statistical properties of single sodium channels.}
\newblock Journal of General Physiology. 1984;84(4):505--534.
\newblock doi:{10.1085/jgp.84.4.505}.

\bibitem{Clancy1999}
Clancy CE, Rudy Y.
\newblock Linking a genetic defect to its cellular phenotype in a cardiac
  arrhythmia.
\newblock Nature. 1999;400(6744):566--569.
\newblock doi:{10.1038/23034}.

\bibitem{Schwiening2012-dr}
Schwiening CJ.
\newblock A brief historical perspective: Hodgkin and Huxley.
\newblock J Physiol. 2012;590(11):2571--2575.

\bibitem{Stein2008-na}
Stein RB.
\newblock Some models of neuronal variability.
\newblock Biophys J. 2008;7(1):37--68.

\bibitem{FENG2001955}
Feng J.
\newblock Is the integrate-and-fire model good enough?--review.
\newblock Neural Networks. 2001;14(6):955--975.
\newblock doi:{https://doi.org/10.1016/S0893-6080(01)00074-0}.

\bibitem{Gerstner2002}
Gerstner W, Kistler WM. Spiking neuron models : single neurons, populations,
  plasticity; 2002.
\newblock Available from: \url{https://doi.org/10.1017/CBO9780511815706}.

\bibitem{Lansky2006}
Lansky P, Sanda P, He J.
\newblock The parameters of the stochastic leaky integrate-and-fire neuronal
  model.
\newblock Journal of Computational Neuroscience. 2006;21(2):211--223.
\newblock doi:{10.1007/s10827-006-8527-6}.

\bibitem{10.3389/fninf.2019.00071}
Igarashi J, Yamaura H, Yamazaki T.
\newblock Large-Scale Simulation of a Layered Cortical Sheet of Spiking Network
  Model Using a Tile Partitioning Method.
\newblock Frontiers in Neuroinformatics. 2019;13.
\newblock doi:{10.3389/fninf.2019.00071}.

\bibitem{IGARASHI2011950}
Igarashi J, Shouno O, Fukai T, Tsujino H.
\newblock Real-time simulation of a spiking neural network model of the basal
  ganglia circuitry using general purpose computing on graphics processing
  units.
\newblock Neural Networks. 2011;24(9):950--960.
\newblock doi:{https://doi.org/10.1016/j.neunet.2011.06.008}.

\bibitem{10.1162/neco.1993.5.2.200}
Srinivasan R, Chiel HJ.
\newblock {Fast Calculation of Synaptic Conductances}.
\newblock Neural Computation. 1993;5(2):200--204.
\newblock doi:{10.1162/neco.1993.5.2.200}.

\bibitem{doi:10.1073/pnas.1821227116}
Hendricks WD, Westbrook GL, Schnell E.
\newblock Early detonation by sprouted mossy fibers enables aberrant dentate
  network activity.
\newblock Proceedings of the National Academy of Sciences.
  2019;116(22):10994--10999.
\newblock doi:{10.1073/pnas.1821227116}.

\bibitem{10.3389/neuro.10.026.2009}
Humphries M, Lepora N, Wood R, Gurney K.
\newblock Capturing dopaminergic modulation and bimodal membrane behaviour of
  striatal medium spiny neurons in accurate, reduced models.
\newblock Frontiers in Computational Neuroscience. 2009;3.
\newblock doi:{10.3389/neuro.10.026.2009}.

\bibitem{brauer2009carl}
Brauer W, Reisig W.
\newblock Carl adam Petri and “Petri nets”.
\newblock Fundamental concepts in computer science. 2009;3(5):129--139.

\bibitem{petri1966communication}
Petri CA.
\newblock Communication with automata: Volume 1 supplement 1.
\newblock DTIC Document. 1966;.

\bibitem{10.1007/978-3-319-07674-4_51}
Thong WJ, Ameedeen MA.
\newblock A Survey of Petri Net Tools.
\newblock In: Sulaiman HA, Othman MA, Othman MFI, Rahim YA, Pee NC, editors.
  Advanced Computer and Communication Engineering Technology. Cham: Springer
  International Publishing; 2015. p. 537--551.

\bibitem{Fisher_Tippett_1928}
Fisher RA, Tippett LHC.
\newblock Limiting forms of the frequency distribution of the largest or
  smallest member of a sample.
\newblock Mathematical Proceedings of the Cambridge Philosophical Society.
  1928;24(2):180^^e2^^80^^93190.
\newblock doi:{10.1017/S0305004100015681}.

\bibitem{Frechet_Maurice_Sur_1928}
Frech\'{e}t M. type [; 1928].

\bibitem{Gnedenko}
Gnedenko B.
\newblock Sur La Distribution Limite Du Terme Maximum D'Une Serie Aleatoire.
\newblock Annals of Mathematics. 1943;44(3):423--453.

\bibitem{Taylor1985}
Taylor AE.
\newblock A study of Maurice Fr{\'e}chet: II. Mainly about his work on general
  topology, 1909--1928.
\newblock Archive for History of Exact Sciences. 1985;34(4):279--380.
\newblock doi:{10.1007/BF00411640}.

\bibitem{doi:10.1152/jn.1999.82.3.1352}
Wierenga CJ, Wadman WJ.
\newblock Miniature Inhibitory Postsynaptic Currents in CA1 Pyramidal Neurons
  After Kindling Epileptogenesis.
\newblock Journal of Neurophysiology. 1999;82(3):1352--1362.
\newblock doi:{10.1152/jn.1999.82.3.1352}.

\bibitem{doi:10.1073/pnas.1209798109}
Trigo FF, Sakaba T, Ogden D, Marty A.
\newblock Readily releasable pool of synaptic vesicles measured at single
  synaptic contacts.
\newblock Proceedings of the National Academy of Sciences.
  2012;109(44):18138--18143.
\newblock doi:{10.1073/pnas.1209798109}.

\bibitem{Saveliev2015}
Saveliev A, Khuzakhmetova V, Samigullin D, Skorinkin A, Kovyazina I, Nikolsky
  E, et~al.
\newblock Bayesian analysis of the kinetics of quantal transmitter secretion at
  the neuromuscular junction.
\newblock Journal of Computational Neuroscience. 2015;39(2):119--129.
\newblock doi:{10.1007/s10827-015-0567-3}.

\bibitem{10.1371/journal.pone.0096485}
Knoblauch A, K^^c3^^b6rner E, K^^c3^^b6rner U, Sommer FT.
\newblock Structural Synaptic Plasticity Has High Memory Capacity and Can
  Explain Graded Amnesia, Catastrophic Forgetting, and the Spacing Effect.
\newblock PLOS ONE. 2014;9(5):1--19.
\newblock doi:{10.1371/journal.pone.0096485}.

\bibitem{10.1172/JCI169064}
Kauwe G, Pareja-Navarro KA, Yao L, Chen JH, Wong I, Saloner R, et~al.
\newblock KIBRA repairs synaptic plasticity and promotes resilience to
  tauopathy-related memory loss.
\newblock The Journal of Clinical Investigation. 2024;134(3).
\newblock doi:{10.1172/JCI169064}.

\bibitem{10.1371/journal.pcbi.1002836}
van Rossum MCW, Shippi M, Barrett AB.
\newblock Soft-bound Synaptic Plasticity Increases Storage Capacity.
\newblock PLOS Computational Biology. 2012;8(12):1--11.
\newblock doi:{10.1371/journal.pcbi.1002836}.

\bibitem{60304975-67e2-38df-86b7-1fdac19c6683}
Hosking JRM, Wallis JR, Wood EF.
\newblock Estimation of the Generalized Extreme-Value Distribution by the
  Method of Probability-Weighted Moments.
\newblock Technometrics. 1985;27(3):251--261.

\bibitem{Singh1998}
Singh VP.
\newblock In: Generalized Extreme Value Distribution. Dordrecht: Springer
  Netherlands; 1998. p. 169--183.

\bibitem{Abdulali2022}
Abdulali BAA, Abu~Bakar MA, Ibrahim K, Mohd~Ariff N.
\newblock Extreme Value Distributions: An Overview of Estimation and
  Simulation.
\newblock Journal of Probability and Statistics. 2022;2022:5449751.
\newblock doi:{10.1155/2022/5449751}.

\bibitem{AMIN2021123}
Amin MT, Khan F, Ahmed S, Imtiaz S.
\newblock Risk-based fault detection and diagnosis for nonlinear and
  non-Gaussian process systems using R-vine copula.
\newblock Process Safety and Environmental Protection. 2021;150:123--136.
\newblock doi:{https://doi.org/10.1016/j.psep.2021.04.010}.

\bibitem{7408339}
Hubert S, Baur F, Delgado A, Helmers T, R^^c3^^a4biger N.
\newblock Simulation modeling of bottling line water demand levels using
  reference nets and stochastic models.
\newblock In: 2015 Winter Simulation Conference (WSC); 2015. p. 2272--2282.

\bibitem{7500642}
Geng F, Dubos GF, Saleh JH.
\newblock Spacecraft obsolescence: Modeling, value analysis, and implications
  for design and acquisition.
\newblock In: 2016 IEEE Aerospace Conference; 2016. p. 1--13.

\bibitem{doi:10.1177/0954409713481725}
Rama D, Andrews JD.
\newblock A reliability analysis of railway switches.
\newblock Proceedings of the Institution of Mechanical Engineers, Part F:
  Journal of Rail and Rapid Transit. 2013;227(4):344--363.
\newblock doi:{10.1177/0954409713481725}.

\bibitem{NAOI2022400a}
Naoi T, Kagawa Y, Nagino K, Niwa S, Hayashi K.
\newblock Extreme value analysis of the velocity of axonal transport by kinesin
  and dynein.
\newblock Biophysical Journal. 2022;121(3, Supplement 1):400a.
\newblock doi:{https://doi.org/10.1016/j.bpj.2021.11.766}.

\bibitem{sbbd_estendido}
Fonseca A, Porto F, Ferro M, Ogasawara E, Bezerra E.
\newblock Analysis of precipitation data in Rio de Janeiro city using Extreme
  Value Theory.
\newblock In: Anais Estendidos do XXXVII Simp^^c3^^b3sio Brasileiro de Bancos
  de Dados. Porto Alegre, RS, Brasil: SBC; 2022. p. 193--198.

\bibitem{Artha_2019}
Artha AA, Sofro A.
\newblock Rainfall Prediction in East Java Using Spatial Extreme Value Theory.
\newblock Journal of Physics: Conference Series. 2019;1417(1):012010.
\newblock doi:{10.1088/1742-6596/1417/1/012010}.

\bibitem{doi:10.1080/23737484.2020.1789901}
Guermah T, Rassoul A.
\newblock Study of extreme rainfalls using extreme value theory (case study:
  Khemis-Miliana region, Algeria).
\newblock Communications in Statistics: Case Studies, Data Analysis and
  Applications. 2020;6(3):364--379.
\newblock doi:{10.1080/23737484.2020.1789901}.

\bibitem{einmahl2008records}
Einmahl JH, Magnus JR.
\newblock Records in athletics through extreme-value theory.
\newblock Journal of the American Statistical Association.
  2008;103(484):1382--1391.

\bibitem{Einmahl2011}
Einmahl JHJ, Smeets SGWR.
\newblock Ultimate 100-m world records through extreme-value theory.
\newblock Statistica Neerlandica. 2011;65(1):32--42.
\newblock doi:{https://doi.org/10.1111/j.1467-9574.2010.00470.x}.

\bibitem{ArderiudeFondeville}
Arderiu A, de~Fondeville R.
\newblock Influence of advanced footwear technology on sub-2 hour marathon and
  other top running performances.
\newblock Journal of Quantitative Analysis in Sports. 2022;18(1):73--86.
\newblock doi:{doi:10.1515/jqas-2021-0043}.

\bibitem{PhysRevE.98.042102}
Tsuzuki S, Yanagisawa D, Nishinari K.
\newblock Effect of walking distance on a queuing system of a totally
  asymmetric simple exclusion process equipped with functions of site
  assignments.
\newblock Phys Rev E. 2018;98:042102.
\newblock doi:{10.1103/PhysRevE.98.042102}.

\bibitem{PhysRevE.80.051119}
Arita C.
\newblock Queueing process with excluded-volume effect.
\newblock Phys Rev E. 2009;80:051119.
\newblock doi:{10.1103/PhysRevE.80.051119}.

\bibitem{PhysRevE.84.051127}
Arita C, Schadschneider A.
\newblock Exact dynamical state of the exclusive queueing process with
  deterministic hopping.
\newblock Phys Rev E. 2011;84:051127.
\newblock doi:{10.1103/PhysRevE.84.051127}.

\bibitem{PhysRevE.83.051128}
Arita C, Schadschneider A.
\newblock Dynamical analysis of the exclusive queueing process.
\newblock Phys Rev E. 2011;83:051128.
\newblock doi:{10.1103/PhysRevE.83.051128}.

\bibitem{Simons-Weidenmaier2006}
Simons-Weidenmaier NS, Weber M, Plappert CF, Pilz PK, Schmid S.
\newblock Synaptic depression and short-term habituation are located in the
  sensory part of the mammalian startle pathway.
\newblock BMC Neuroscience. 2006;7(1):38.
\newblock doi:{10.1186/1471-2202-7-38}.

\bibitem{cinlar2013introduction}
Cinlar E.
\newblock Introduction to stochastic processes.
\newblock Courier Corporation; 2013.

\bibitem{Sekirnjak2002}
Sekirnjak C, du~Lac S.
\newblock Intrinsic firing dynamics of vestibular nucleus neurons.
\newblock J Neurosci. 2002;22(6):2083--2095.

\bibitem{doi:10.1152/jn.00510.2004}
Geisler C, Brunel N, Wang XJ.
\newblock Contributions of Intrinsic Membrane Dynamics to Fast Network
  Oscillations With Irregular Neuronal Discharges.
\newblock Journal of Neurophysiology. 2005;94(6):4344--4361.
\newblock doi:{10.1152/jn.00510.2004}.

\bibitem{doi:10.1152/jn.1997.77.2.537}
Negro CAD, Chandler SH.
\newblock Physiological and Theoretical Analysis of K+ Currents Controlling
  Discharge in Neonatal Rat Mesencephalic Trigeminal Neurons.
\newblock Journal of Neurophysiology. 1997;77(2):537--553.
\newblock doi:{10.1152/jn.1997.77.2.537}.

\bibitem{doi:10.1137/1.9780898717570}
Vogel C.
\newblock Computational Methods for Inverse Problems.
\newblock The SIAM series on Frontiers in Applied Mathematics; 2002.

\bibitem{Lecuyer2011}
Devroye L.
\newblock Non-uniform Random Variate Generation.
\newblock Springer-Verlag; 1986.

\bibitem{Boudkkazi2007-lk}
Boudkkazi S, Carlier E, Ankri N, Caillard O, Giraud P, Fronzaroli-Molinieres L,
  et~al.
\newblock Release-dependent variations in synaptic latency: a putative code for
  short- and long-term synaptic dynamics.
\newblock Neuron. 2007;56(6):1048--1060.

\bibitem{Boudkkazi2011-pn}
Boudkkazi S, Fronzaroli-Molinieres L, Debanne D.
\newblock Presynaptic action potential waveform determines cortical synaptic
  latency.
\newblock J Physiol. 2011;589(Pt 5):1117--1131.

\bibitem{Boudkkazi2024-bq}
Boudkkazi S, Debanne D.
\newblock Enhanced Release Probability without Changes in Synaptic Delay during
  {Analogue-Digital} Facilitation.
\newblock Cells. 2024;13(7).

\bibitem{Izhikevich2006-ll}
Izhikevich EM.
\newblock Polychronization: computation with spikes.
\newblock Neural Comput. 2006;18(2):245--282.

\bibitem{Chowdhury_2019}
Chowdhury SN, Majhi S, Ozer M, Ghosh D, Perc M.
\newblock Synchronization to extreme events in moving agents.
\newblock New Journal of Physics. 2019;21(7):073048.
\newblock doi:{10.1088/1367-2630/ab2a1f}.

\bibitem{10.1063/5.0131133}
Gao M, Zhao Y, Wang Z, Wang Y.
\newblock {A modified extreme event-based synchronicity measure for climate
  time series}.
\newblock Chaos: An Interdisciplinary Journal of Nonlinear Science.
  2023;33(2):023105.
\newblock doi:{10.1063/5.0131133}.

\end{thebibliography}

\end{document}